\documentclass[12,preprint]{emulateapj}

\shorttitle{COSMOS AGN}
\shortauthors{Trump et~al.}

\begin{document}

\title{Magellan Spectroscopy of AGN Candidates in the COSMOS
Field\altaffilmark{\ref{cosmos}}}

\author{
  Jonathan R. Trump,\altaffilmark{\ref{Arizona}}
  Chris D. Impey,\altaffilmark{\ref{Arizona}}
  Patrick J. McCarthy,\altaffilmark{\ref{Carnegie}}
  Martin Elvis,\altaffilmark{\ref{CfA}}
  John P. Huchra,\altaffilmark{\ref{CfA}}
  Marcella Brusa,\altaffilmark{\ref{Max Planck}}
  Gunther Hasinger,\altaffilmark{\ref{Max Planck}}
  Eva Schinnerer,\altaffilmark{\ref{Max Planck astro}}
  Peter Capak,\altaffilmark{\ref{Caltech}}
  Simon J. Lilly,\altaffilmark{\ref{Zurich}}
  and Nick Z. Scoville\altaffilmark{\ref{Caltech}}
}

\altaffiltext{1}{
  Based on observations with the NASA/ESA \emph{Hubble Space
  Telescope}, obtained at the Space Telescope Science Institute, which
  is operated by AURA Inc, under NASA contract NAS 5-26555; and based
  on data collected at the Magellan Telescope, which is operated by
  the Carnegie Observatories.
\label{cosmos}}

\altaffiltext{2}{
   Steward Observatory, University of Arizona, 933 North Cherry
   Avenue, Tucson, AZ 85721
\label{Arizona}}

\altaffiltext{3}{
  Observatories of the Carnegie Institute of Washington, Santa Barbara
  Street, Pasadena, CA 91101
\label{Carnegie}}

\altaffiltext{4}{
  Harvard-Smithsonian Center for Astrophysics, 60 Garden Street,
  Cambridge, MA 02138
\label{CfA}}

\altaffiltext{5}{
  Max Planck-Institut f\"ur Extraterrestrische Physik,
  Giessenbachstrasse 1, D-85748 Garching, Germany
\label{Max Planck}}

\altaffiltext{6}{
  Max Planck Institut f\"ur Astronomie, K\"onigstuhl 17, D-69117
  Heidelberg, Germany
\label{Max Planck astro}}

\altaffiltext{7}{
  Department of Physics, ETH Zurich, CH-8093 Zurich, Switzerland
\label{Zurich}}

\altaffiltext{8}{
  California Institute of Technology, MC 105-24, 1200 East California
  Boulevard, Pasadena, CA 91125
\label{Caltech}}

\newcommand{\MgII}{\hbox{{\rm Mg}\kern 0.1em{\sc ii}}}
\newcommand{\CIII}{\hbox{{\rm C}\kern 0.1em{\sc iii}]}}
\newcommand{\OII}{\hbox{{\rm [O}\kern 0.1em{\sc ii}]}}

\begin{abstract}

We present spectroscopic redshifts for the first 466 X-ray and
radio-selected AGN targets in the 2 deg$^2$ COSMOS field.  Spectra
were obtained with the IMACS instrument on the Magellan (Baade)
telescope, using the nod-and-shuffle technique.  We identify a variety
of Type 1 and Type 2 AGN, as well as red galaxies with no emission
lines.  Our redshift yield is 72\% down to $i_{\rm AB}=24$, although
the yield is $>90\%$ for $i_{AB}<22$.  We expect the completeness to
increase as the survey continues.  When our survey is complete and
additional redshifts from the zCOSMOS project are included, we
anticipate $\sim 1100$ AGN with redshifts over the entire COSMOS
field.  Our redshift survey is consistent with an obscured AGN
population that peaks at $z \sim 0.7$, although further work is
necessary to disentangle the selection effects.

\end{abstract}


\section{Introduction}

The Cosmic Evolution Survey \citep[COSMOS, ][]{cosmos} is an HST
Treasury project to fully image a 2 deg$^2$ equatorial field.  The 590
orbits of HST ACS $i$-band observations have been supplemented by
observations at wavelengths from X-ray to radio and a major galaxy
redshift survey \citep[zCOSMOS, ][]{zcosmos} carried out with
VLT/VIMOS.  The details of the COSMOS AGN survey are found in a
companion paper \citep{cosmosagn}.  Here we present the first X-ray
and radio-selected active galactic nuclei (AGN) candidates observed
with the IMACS instrument on the Magellan (Baade) telescope.

X-ray observations provide the most efficient method to find Type 1,
Type 2, and particularly obscured AGN.  The XMM-Newton observations of
the COSMOS field are expected to reach an AGN surface density of
$\sim$1000 deg$^2$.  The current COSMOS X-ray catalog is presented by
\citet{cosmosxray} and has a 0.5-2 keV flux limit of $1 \times
10^{-15}~{\rm erg~cm^{-2}~s^{-1}}$ and a 2-10 keV flux limit of $3.3
\times 10^{-15}~{\rm erg~cm^{-2}~s^{-1}}$.  The identification of
optical counterparts, based on the ``likelihood ratio'' technique, is
presented by \citet{brusa}.  The X-ray selected targets for our IMACS
survey were the $i_{\rm AB}<24$ optical counterparts of \citet{brusa}
that were X-ray point sources with detection in either the 0.5-2 keV
or 2-10 keV bands, available at the time of our IMACS observations.
Multiple X-ray observations over most of the COSMOS field mitigate the
effects of vignetting in the outer region of the XMM-Newton field of
view.  The edges of the COSMOS field, however, are observed only once
by XMM-Newton and so our observations in these regions must sample a
lower density of X-ray selected AGN candidates.

Radio-selected AGN candidates were our second-highest priority targets
for IMACS observations.  The COSMOS VLA survey is described by
\citet{cosmosradio}; we use a preliminary VLA catalog with a $4\sigma$
flux limit of 0.1-0.4 mJy at 1.4 GHz and full coverage across the
COSMOS field.  Approximately 20\% of the radio selected AGN candidates
overlapped with the X-ray sample.  We observed only radio sources with
radio peak flux $S/N\ge 4$ and unambiguous optical counterparts within
1'' of the radio peak of magnitude $i_{\rm AB}<24$.

In \S 2 we present the details of our observing strategy and set-up,
as well as the reduction and calibration of the observations.  We
present the classifications and redshifts of our targets in \S 3,
along with estimates of our completeness and other properties of the
sample.  We summarize our results in \S 4 and discuss our timeline for
completing the survey.

\section{Observations}

\subsection{Instrumental Setup}

Our observations were taken with the Inamori Magellan Areal Camera and
Spectrograph \citep[IMACS,][]{imacs}.  The field of view of the IMACS
camera is $22{\arcmin}30{\arcsec} \times 21{\arcmin}10{\arcsec}$, so
that a tiling of 16 IMACS pointings will cover the entire 2 deg$^2$
COSMOS field.  The tiling that we adopted is shown in Figure 1.
Henceforth we will refer to each field by the number designation shown
in this figure.  In this paper we present the 7 pointings observed
during the nights of January 16-19, February 8-10 and February 12-15,
2005.  These pointings (designated by 6, 7, 10, 11, 12, 15, and 16)
are shown as shaded circles in Figure 1.  At the time of our
observations, the entire field had been uniformly observed by the VLA,
but the XMM observations were not complete.  The available X-ray and
radio AGN candidates are overplotted in Figure 1 as crosses and
diamonds, respectively.  The 7 observed fields presented here had the
greatest surface densities of available X-ray targets.

In all fields, the X-ray and radio candidates were given the highest
priority except for a rare set of ``must-have'' objects.  There were
$<20$ ``must-have'' targets in each pointing, and their inclusion
eliminated no more than 5 X-ray and radio targets from each IMACS
mask.  On average, we were able to target 75\% of X-ray candidates and
73\% of the radio candidates (with $\sim$20\% overlap between radio
and X-ray targets).  Most of the objects not targeted for IMACS
observations, in addition to some that are too faint for IMACS, will
be observed with VLT/VIMOS as part of the zCOSMOS galaxy redshift
survey \citep{zcosmos}.

We observed with the ``short'' f/2 camera and the 200 line grating
centered at 6646\AA, which delivers a $\sim$5 pixel resolution element
of $\sim$10\AA.  All observations were taken with the Moon below the
horizon and airmass in a range of 1 to 1.8, with a mean airmass of
1.3.  The January observations used the OG570 filter for a wavelength
range of 5600-9200\AA, while for the February run we upgraded to the
new 565-920 filter with better throughput and a wavelength range of
5400-9200\AA.

We cut three different masks for each pointing: a ``nod-and-shuffle,''
a ``poor-seeing,'' and a conventional mask.  Because new X-ray targets
became available after the January run, we also cut new masks for the
February observations.  Nod-and-shuffle masks were used in all cases
with seeing $\lesssim 1\arcsec$, which was true for all observations
presented in this paper except field 10, which was partially observed
with a poor-seeing mask and seeing $>1\arcsec$.  Conventional masks
were designed to be used only if the IMACS nod-and-shuffle mode was
not working.  Since our nod-and-shuffle observations operated
smoothly, the conventional masks were not used, and so we omit them
from the discussion.  The nod-and-shuffle and poor-seeing masks are
discussed in detail below.  Each field was observed for no more than
3600 seconds at a time before re-aligning the telescope.  The total
exposure times for each pointing are listed in Table 1, along with
totals for the first season of observing and projections for the
coverage of the entire COSMOS field.

The nod-and-shuffle masks were designed for the ideal case of seeing
$\lesssim 1\arcsec$.  The nod-and-shuffle technique in spectroscopic
observations has been shown to allow sky subtraction and fringe
removal an order of magnitude more precisely than conventional methods
(e.g., \citealp{gemini}).  \citet{nodshuffle} describe the principles
of the nod-and-shuffle technique, and our specific nod-and-shuffle
strategy is detailed in Appendix 1 of \citet{gemini}.  In the nod and
shuffle masks we reserved $11\arcsec \times 1\arcsec$ ($55 \times 5$
pixels) for each object, but only $5{\farcs}4 \times 1\arcsec$ was cut
into a slit, so that an extra adjacent $5{\farcs}6$ was reserved.  We
observed each object for 60 seconds, then closed the shutter, nodded
the telescope by 9 pixels ($1{\farcs}8$), and shuffled the charge to
the reserved ``uncut region''.  We then observed for 60 seconds in the
new position so that the sky was observed on the same pixels as the
original target.  Then the shutter closed, the charge was shuffled,
and the telescope was nodded back to the original position and the
cycle repeated (typically 15 times).  Our slit width and nod distance
were appropriate for the $\lesssim 1\arcsec$ seeing of our
nod-and-shuffle observations.

The poor-seeing masks had larger $12\arcsec \times 1{\farcs}25$ slits
and a magnitude cut of $i_{\rm AB} < 23$ designed for seeing
$>1\arcsec$ and/or thin cloud cover.  Field 10 is the only pointing in
which we present poor-seeing mode observations.  While the sky
subtraction is inferior to that of the nod-and-shuffle, the shallower
magnitude cut allows us to extract spectra and measure redshifts with
roughly the same efficiency as in the nod-and-shuffle observations.

In Table 1 we show the number of X-ray and radio targets in each mask.
Fields 7 and 10 were observed with different January and February
masks, and the numbers of objects and exposure times listed in Table 1
are the combined totals of unique targets and the combined exposure
times.  About $30\%$ of the X-ray and radio targets in fields 7 and 10
were observed in only January or February.

\subsection{Data Reduction}

We used the publicly available Carnegie Observatories System for
MultiObject Spectroscopy (with coincidentally the same acronym COSMOS,
written by A. Oemler) to extract and sky-subtract individual 2D linear
spectra.  We combined the nodded positions in the nod-and-shuffle data
and co-added and cosmic ray subtracted the individual observations of
each pointing.  The spectra were wavelength and flux calibrated using
the IDL {\tt ispec2d} package \citep{ispec}.  Wavelength calibration
was performed using an arc lamp exposure in each slit.  While flux
calibration used only a single standard star at the center of the
IMACS detector, we estimate that vignetting has $<10\%$ effect on the
spectral shape or throughput across the field.  We wrote our own IDL
software to extract 1D spectra from the individual 2D frames.

IMACS spectra are contaminated or compromised from several major
sources, including 0th and 2nd order lines from other spectra, bad
pixels and columns, chip gaps, poorly machined slits, and cosmic rays
missed in the coadding stage.  To eliminate these artifacts, we
generated masks for all spectra by visual inspection of the calibrated
1D and 2D data.  The nod-and-shuffle 2D data were especially useful
for artifact rejection: with two nod-separated spectra, any feature
appearing in only one of the nod positions is clearly an artifact.

Data from the January and February runs in fields 7 and 10 were only
combined when the fully reduced 1D spectrum from one mask was too poor
to find a reliable redshift.  The unmasked 1D spectra were combined,
weighting by exposure time (half exposure time for the poor-seeing
observations, based on the signal-to-noise impact of increased image
size).  A total of 17 objects used data combined from the January and
February runs, and 3 of these gained new redshifts after the
combinations.  Objects in fields 7 and 10 with a well-exposed spectra
and a reliable redshift in both the January and February runs had
redshifts that matched within the errors.

\section{Results}

\subsection{Classification and Redshift Determination}

We used three composite spectra from the Sloan Digital Sky Survey
\citep[SDSS;][]{sdss} as templates for the classification and redshift
determination of our objects: a Type 1 AGN composite from
\citet{q1template}, a Type 2 AGN composite from \citet{q2template},
and a red galaxy composite from \citet{etemplate}.  The three template
spectra are shown in Figure 2.  Objects showing a mix of Type 2 AGN
narrow emission lines and red galaxy continuum shape and absorption
features were classified as hybrid objects.

To calculate redshifts we used a cross-correlation redshift IDL
algorithm in the publicly available {\tt idlspec2d} package written by
David Schlegel.  This algorithm used our visually-classified template
to find a best-fit redshift and its associated error.  All masked-out
regions were ignored in the redshift determination.  Note that the
error returned is probably underestimated for objects with lines
shifted from the rest frame with respect to each other, as is often
the case in AGN \citep{lineshifts}.  We manually assigned redshift
errors for a small fraction of objects where the cross-correlation
algorithm was unable to find a best-fit redshift.

Each object was assigned a redshift confidence according to the
ability of the redshifted template to fit the emission lines,
absorption lines, and continuum of the object spectrum.  If at least
two emission or absorption lines were fit well, or if at least one
line and the minor continuum features were fit well, the redshift was
considered unambiguous and assigned $z_{\rm conf}=1$.  Six objects
with $z_{\rm conf}=1$ redshifts are shown in Figures 3 and 4.  If only
one line could be fit, or if the redshift came strictly from a
well-fit continuum shape over the entire spectral range, the object
was assigned $z_{\rm conf}=2$.  Two $z_{\rm conf}=2$ objects are shown
in Figures 3 (bottom) and 4 (second from bottom).  If the
signal-to-noise of the object spectrum was too low for a redshift to
be determined, it was assigned $z_{\rm conf}=~?$.  Of our X-ray
targets, 60\% were assigned $z_{\rm conf}=1$, 12\% were $z_{\rm
conf}=2$, and 28\% were $z_{\rm conf}=~?$ or undetermined.  The radio
targets had 63\% with $z_{\rm conf}=1$, 10\% with $z_{\rm conf}=2$,
and 26\% with $z_{\rm conf}=~?$ or undetermined.

All of the objects observed in our sample are presented in Table 2.
The classifications are as follows: ``q1'' for Type 1 AGN, ``q2'' for
Type 2 AGN, ``e'' for red galaxy, ``q2e'' for Type 2 AGN and red
galaxy hybrids, and ``mstar'' for M-type stars.  We designate
questionable classifications with a question mark: objects with blue
continua but no obvious emission lines are listed as ``q?''  and
objects with red continua and no emission or absorption lines are
listed as ``e?''.  Over all of our observations, 51\% of the
classified X-ray targets were designated ``q1,'' 33\% were ``q2'' or
``q2e,'' and 17\% were ``e.''  These classification fractions roughly
agree with other wide-area X-ray surveys such as those of
\citet{survey1}, \citet{survey2}, and \citet{survey3}.  For the radio
targets, 2\% were classified as ``q1,'' 64\% were ``q2'' or ``q2e,''
and 33\% were ``e.''  Objects with a question mark under ``Type'' in
Table 2 have too low signal-to-noise to venture a classification,
although many of these objects are unlikely to be Type 1 or 2 AGN.
Some objects have classifications without redshifts, although the
reverse is not true.  We summarize our efficiencies, from targeting to
redshifts, in Table 3.

Many of the objects with red galaxy spectra are probably optically
obscured Type 2 AGN because of their X-ray and radio
emission. However, other large radio surveys of AGN (e.g.,
\citealp{best}, \citealp{sadler}) suggest that a significant fraction
of our radio selected ``Type 2 AGN'' are actually star-forming
galaxies.  We make no distinction between Type 2 AGN and emission-line
galaxies: all objects with narrow emission lines are classified as
``q2'' or ``q2e'' objects.  We will fully distinguish between the
star-forming and AGN-dominated galaxies in future work
(\citealp{linemeasure}, \citealp{radiotype2}).

\subsection{Redshift Completeness}

To use our spectroscopic sample for science, it is necessary to
understand our completeness in classifying and assigning redshifts.
Our completeness ultimately depends on spectral signal-to-noise, but
it is more useful to understand completeness as a function of
magnitude.  The spectral signal-to-noise per pixel and target $i_{\rm
AB}$ magnitudes for our different classified types are shown in Figure
5.  In general, the signal-to-noise is correlated with the $i_{\rm
AB}$ magnitude, consistent with the goal of a uniform spectroscopic
survey.  Outlying objects were visually inspected and found to have
inaccurate spectra caused by poorly cut or misaligned slits, or by
extreme contamination from artifacts.  Figure 6 shows our redshift
yield with magnitude and signal-to-noise.  Our overall redshift yield
drops significantly for objects of $S/N \lesssim 2.5$, corresponding
to $i_{\rm AB} \gtrsim 22$.  However, we might expect our redshift
yields to be better for Type 1 and 2 AGN because they have prominent
emission lines.

Our classification completeness by type is shown in Figure 7.  The
classification completeness corresponds roughly to the redshift
completeness, although more objects are classified than assigned
$z_{\rm conf}=1$ redshifts.  The number of unclassified objects (the
region labeled ``?'') increases, and our overall completeness
decreases, for $i_{\rm AB} \gtrsim 22$.  But our completeness is not
uniform for all types of objects: the fraction of Type 1 AGN remains
flat to a magnitude bin fainter than the other targets, until $i_{\rm
AB} \gtrsim 23$.  Since Type 2 AGN also have prominent emission lines,
we might expect the same trend as in Type 1 AGN, but this is not the
case.  The decrease in the fraction of Type 2 AGN for $i_{\rm AB}
\gtrsim 21$ is explained by the redshift dependence of our
completeness.

We use Monte Carlo simulations to test the redshift dependence of our
survey's completeness for Type 1 and Type 2 AGN.  We do not simulate
our redshift completeness to ``e'' type objects because the red galaxy
spectra in our sample are well-populated with absorption lines and
their identification should be redshift-independent.  We assume that
the SDSS Type 1 composite spectrum \citep{q1template} and Type 2
composite spectrum \citep{q2template} each have infinite
signal-to-noise, and degrade these spectra with Gaussian-distributed
random noise to artificial values of signal-to-noise.  We then
determined whether or not we would be able to assign a high-confidence
($z_{\rm conf}=1$) redshift for these artificial spectra at various
redshifts (a redshift could not be determined if the emission lines
were smeared out or if the spectrum could not be distinguished from a
different line at another redshift).  The fraction of artificial
spectra with determined redshifts at a given redshift and
signal-to-noise, with different seeds of randomly-added noise, forms
an estimate of our completeness.

We found that our simulated completeness for Type 2 AGN was 90\%
complete to $i_{\rm AB} \sim 23$ ($S/N \approx 1.3$ per pixel) for
$z<1$, with several strong, unambiguous lines.  This is a magnitude
fainter than the level of the average redshift completeness of the
survey.  However, at $z \gtrsim 1$, $\OII~\lambda5007$ is the only one
strong line in our Type 2 AGN spectra and it is difficult to assign a
$z_{\rm conf}=1$ redshift.  Most Type 2 AGN at $z \gtrsim 1$ have
$i_{\rm AB} \gtrsim 21$, so that incompleteness at $z \gtrsim 1$
translates to incompleteness at $i_{\rm AB} \gtrsim 21$, as observed
for Type 2 AGN in Figure 7.

The Type 1 AGN completeness has more complex redshift dependence and
is shown in Figure 8.  We have poorer redshift completeness in the
redshift ranges $1.3 \lesssim z \lesssim 1.4$ and $2.3 \lesssim z
\lesssim 2.45$, where only one line is present ($\MgII$ and $\CIII$,
respectively) and although we can reliably classify as a Type 1 AGN
it is difficult to distinguish between the two redshift ranges.
Without the degeneracies between redshift, our redshifts would be
$>80\%$ complete to $S/N \approx 1.9$ ($i_{\rm AB} \sim 23$).
Because we can generally assign redshifts for Type 1 AGN to a
magnitude fainter than the average survey limit of $i_{\rm AB}
\lesssim 22$, we claim that most of the $i_{\rm AB} \lesssim 23$
unidentified objects in our survey are not Type 1 AGN.

\subsection{Characterizing the Unidentified Targets}

A large fraction ($27\%$) of our spectroscopically observed targets
have spectra too poor for us to venture a classification.  However,
all of our targets have extensive optical broadband photometry as part
of the COSMOS photometric catalog (Capak et~al. 2006).  By comparing
the colors of our unclassified targets to the colors of our classified
targets, we should be able to put constraints on the unclassified
sample.  We find that our classified targets are most strongly
distinguished by their $B-z$ color, displayed against redshift in
Figure 9.  We also find that color separation does not depend on X-ray
versus radio selection; it depends only on the target classification.
Although the colors are most separated at $z \sim 1$, we can use the
$B-z$ color at any redshift to put constraints on the classification
of our poor spectra.

Figure 10 shows our targets with Subaru $B$ and $z$ colors.  Red
galaxies are typically $\sim 3$ magnitudes redder and Type 2 AGN are
$\sim 2$ magnitudes redder than Type 1 AGN.  For $z<23$, our
unclassified targets have colors most consistent with red galaxies and
Type 2 AGN, supporting our simulations which indicate that we are
mostly complete to Type 1 AGN to $i_{\rm AB} \sim 23$ (roughly
similar to $z_{\rm Subaru} \sim 23$ for AGN colors).  For $z>23$,
the unclassified targets span the $B-z$ colors of our different types.

We also use the broadband colors of our Type 1 AGN to attempt to
distinguish between the degenerate redshift ranges of Figure 8.  We
convolve the four Type 1 AGN composites of \citet{q1colors} through
CFHT $u^*$ and Subaru $B$, $V$, $g$, $r$, $i$, and $z$ filters.
Because our Type 1 AGN are not optically selected, their colors at
different redshifts may differ from the simple optically selected Type
1 AGN of \citet{q1colors}.  Therefore to assign a new redshift for a
Type 1 AGN based on its colors, we require evidence from at least two
colors and assume a $z_{\rm conf}=2$ for the new redshift.  Using the
$u^*-B$ and $V-r$ colors, we find 3 Type 1 AGN quasars in the original
redshift range $1.2 \le z \le 1.5$ that have colors more appropriate
for $2.3 \le z \le 2.6$).  We assign these three objects new, higher
redshifts along with $z_{\rm conf}=2$.  One of these redshift-adjusted
Type 1 AGN has its spectrum displayed as the bottom panel of Figure 3.

\subsection{Survey Demographics}

The redshift distribution of our catalog is shown in Figure 11.  The
$z \gtrsim 1$ population is dominated by X-ray selected Type 1 AGN.
The slight statistical excess of Type 1 AGN at $1.1<z<1.3$ might be
affected by the degeneracy between redshifts of $1.1<z<1.4$ and
$2.2<z<2.5$ described in \S 3.3 above.  Although we attempt to resolve
the redshift degeneracy by minor spectral features and broadband
colors, we probably do not completely eliminate the problem.  Only two
radio selected targets are identified as Type 1 AGN, and so we cannot
comment on the radio selected Type 1 AGN population evolution.

There are three effects that contribute to the lack of Type 2 AGN and
red galaxies at $z \gtrsim 1$.  First, Type 2 AGN and red galaxies
have lower optical luminosities than Type 1 AGN, and so are more
difficult to detect at $z \gtrsim 1$.  Our simulations also reveal
that we are incomplete to Type 2 AGN at $z \gtrsim 1$ due to the lack
of strong emission lines in our spectra at these redshifts.  Finally,
recent models of the X-ray luminosity function evolution
(e.g. \citealp{steffen}; \citealp{hasinger}; \citealp{lafranca})
suggest that the distribution of obscured AGN peaks at $z \sim 0.7$,
indicating a physical reason for the lack of obscured AGN at $z
\gtrsim 1$.  Our X-ray Type 2 AGN distribution peaks at $z \sim 0.7$,
consistent with this hypothesis.  However, fully testing the evolution
of the obscured AGN population requires the ability to reliably detect
Type 2 AGN emission lines at $z \gtrsim 1$.  For example, Figure 4 of
\citet{brusa} shows that Type 2 AGN can be detected at higher redshift
by the fainter zCOSMOS survey \citep{zcosmos}.  The radio selected
obscured AGN population is probably better traced by the red galaxies
than the type 2 AGNs, which are contaminated by emission line
galaxies, especially at lower redshifts.  We will disentangle the
radio selected obscured AGN from the star forming galaxies in future
work (\citealp{linemeasure}, \citealp{radiotype2}).

In Figure 12 we show the redshifts of our sample with their target
$i_{\rm AB}$ magnitudes.  The Type 2 AGN and red galaxies appear to
have the same magnitudes at a given redshift, suggesting that Type 2
AGN luminosity is dominated by its host galaxy.  Type 2 AGN and red
galaxies at $z \gtrsim 1$ have $i_{\rm AB} \gtrsim 22$ where our
redshift yield drops.  Type 1 AGN, however, are significantly more
luminous and occupy a distinctly separate region in $z-i_{\rm AB}$
space.  This extends the results of \citet{brusa}, which show the
separate $z-i_{\rm AB}$ regions for X-ray selected Type 1 and 2 AGN.

The absolute $i$-magnitudes of our sample are displayed in Figure 13.
Here we set the (arbitrary) Seyfert/quasar cut at $M_i=-23$.  While
Type 2 and obscured AGN with red galaxy spectra are not often quasars,
we are sensitive to such AGN and identify 10 of these quasars.  We are
also sensitive to the population of Type 1 Seyferts, especially for
$z<1.5$.  We further investigate the luminosities of our AGN in Figure
14, a plot of the X-ray luminosity with redshift.  Our Type 1 AGN are
typically more X-ray luminous than our Type 2 AGN and red galaxies.
The properties of the complete X-ray luminosities, as derived from
spectral analysis, are described in detail by \citet{xrayspec}.

\section{Summary}

The COSMOS AGN survey will provide a large sample of AGN with
bolometric measurements from radio to X-ray and supplementary
observations of their hosts and local environments.  Here we have
presented spectra and redshifts for the first 466 X-ray and
radio-selected AGN targets: we have discovered 86 new Type 1 AGN and
130 new Type 2 AGN with high-confidence redshifts and reliable
classification.  Our overall redshift yield is $72\%$, although we are
$90\%$ complete to objects of $i_{\rm AB} < 22$.  We expect this yield
to increase as refurbishments to IMACS take place.  While the survey
may be affected by redshift-dependent selection effects for $i_{\rm
AB} > 22$, our findings support an obscured AGN population that peaks
at $z \sim 0.7$.  Our observations with IMACS are designed to cover
the entire COSMOS field over three seasons, and a high overall yield
will be obtained thanks to spectra taken by VLT/VIMOS during the
zCOSMOS redshift survey \citep{zcosmos}.  With 7/16 IMACS pointings
successfully observed, we are on schedule to complete our AGN survey
in early 2007.

\acknowledgments

We would like to thank Alan Dressler and the IMACS team for creating
an excellent instrument, as well as telescope operators Hernan
Nu\~{n}ez and Felipe Sanchez and the Las Campanas Observatory staff
for support while observing.  We thank Mike Westover for help with the
January observations.  We also thank the COSMOS team for their work in
creating the catalogs used for selecting our targets.  The HST COSMOS
Treasury program was supported through NASA grant HST-GO-09822.

\begin{center}

\begin{figure}
\plotone{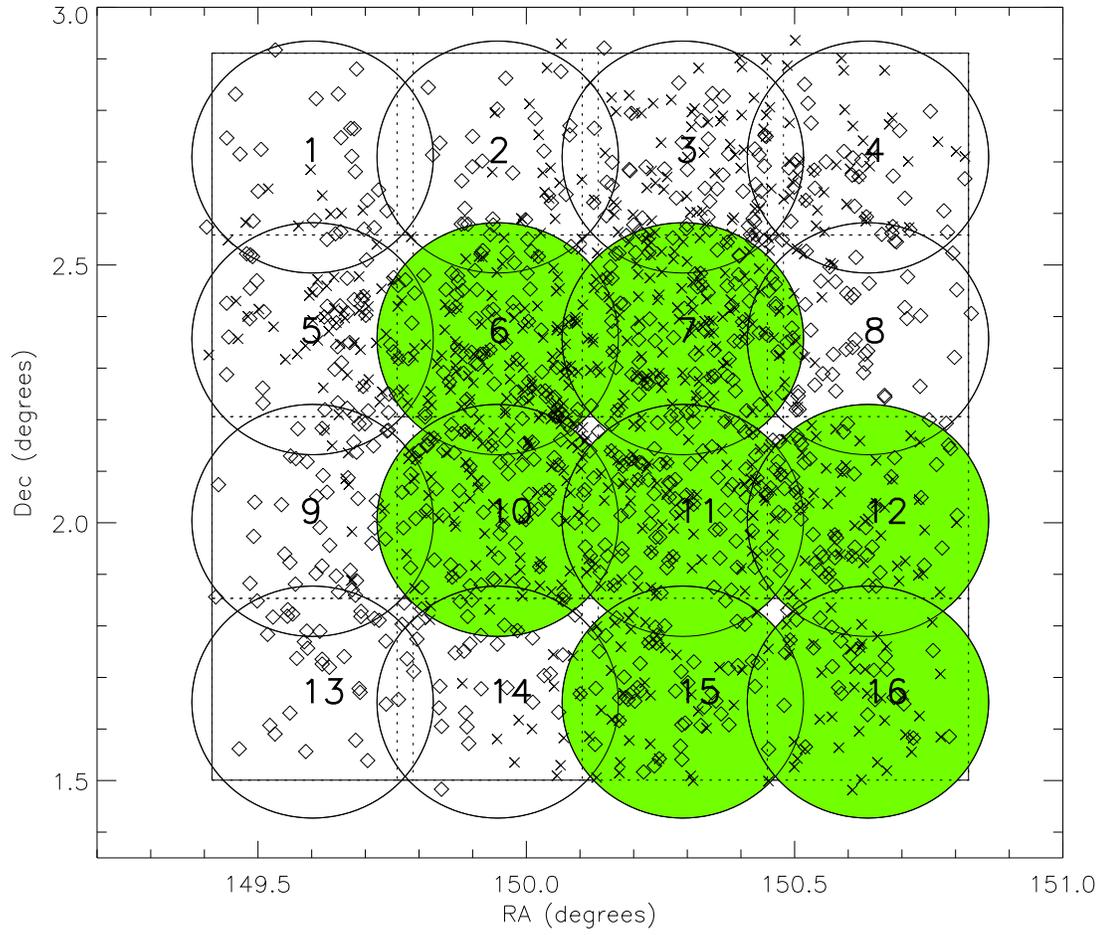}
\figcaption{The sixteen IMACS pointings used to fully cover the COSMOS
area.  The X-ray targets available for our initial observations are
shown as crosses, and radio targets are shown as diamonds.  The seven
shaded fields were chosen to use the regions of maximum X-ray
completeness at the time of our observations, and are the pointings
observed for our initial data set.}
\label{fig:fields}
\end{figure}

\begin{figure}
\plotone{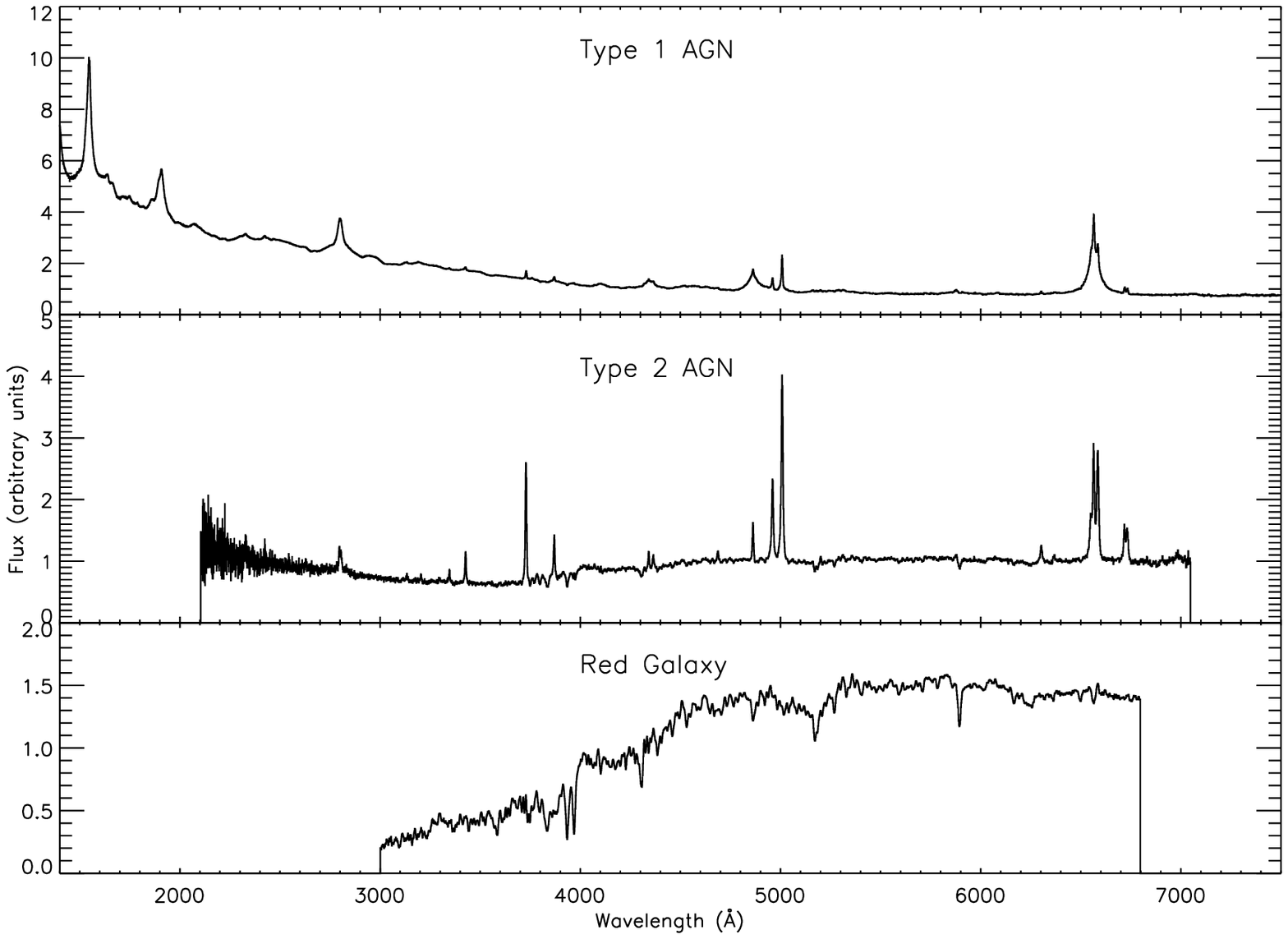}
\figcaption{The three templates used in our classification and
redshift determination scheme.  The Type 1 AGN template is the SDSS
quasar composite of \citet{q1template}, the Type 2 AGN template is the
SDSS Type II AGN composite of \citet{q2template}, and the red galaxy
template is the composite of the SDSS red galaxy sample
\citep{etemplate}.  The wavelength coverages of each template were
within the observed wavelength range for the redshift ranges of the 
different object types in our sample.}
\label{fig:templates}
\end{figure}

\begin{figure}
\plotone{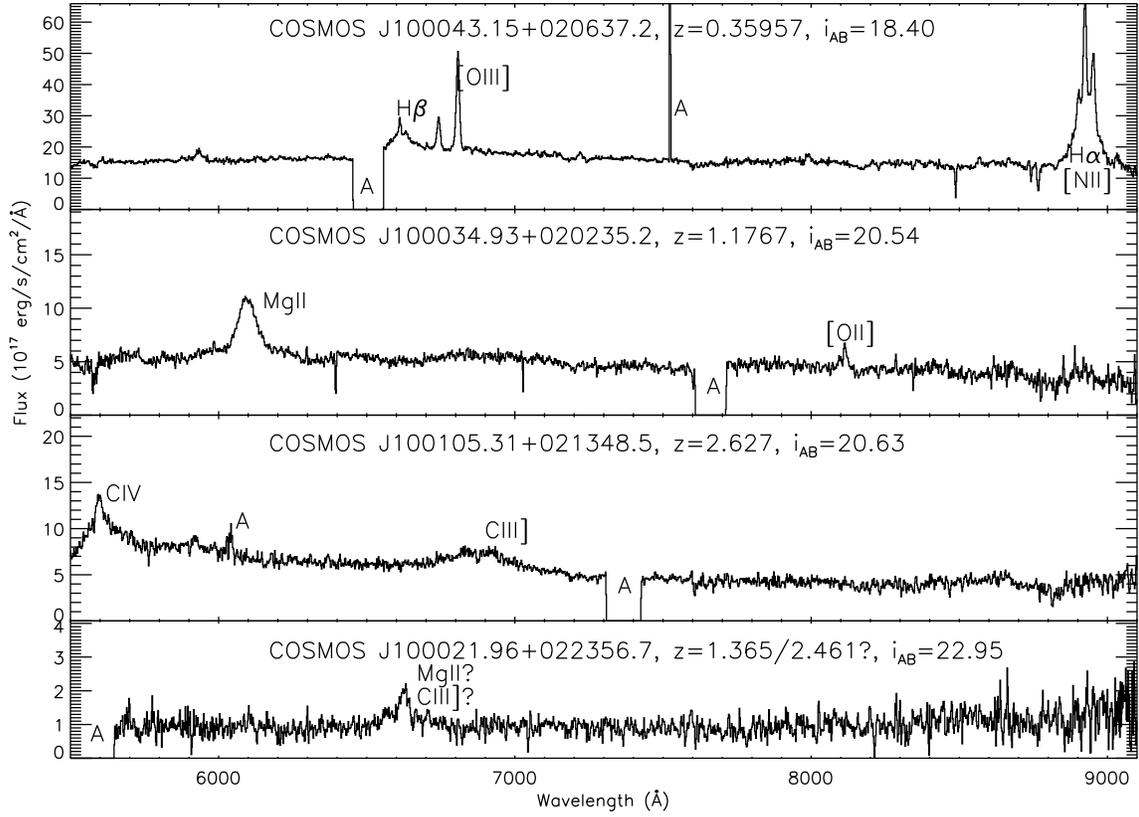}
\figcaption{Four example Type 1 AGN spectra.  Prominent features are
labeled and artifacts are marked by ``A.''  The three objects at top
all have $z_{\rm conf}=1$ and span the range of redshifts sampled by
our observations.  The bottom spectrum, while an unambiguous Type 1
AGN, has roughly equal likelihood of $z=1.365$ as $z=2.461$ and is
therefore assigned $z_{\rm conf}=2$ (its catalog entry is $z=2.461$
because of its broadband colors; see \S 3.3).  Type 1 AGN can be
correctly classified even at low signal-to-noise, but may have an
ambiguous redshift if only one emission line is present.}
\label{fig:type1ex}
\end{figure}

\begin{figure}
\plotone{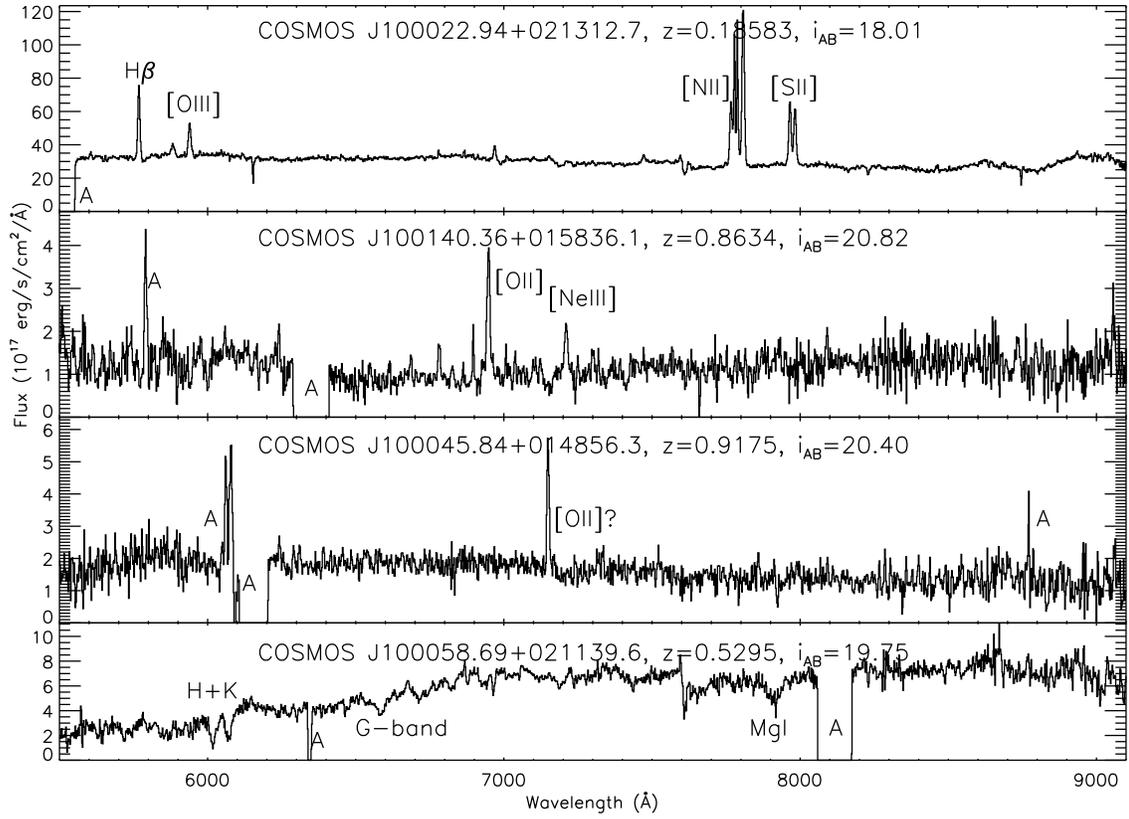}
\figcaption{Three Type 2 AGN spectra and one red galaxy type spectrum.
Prominent features are labeled and artifacts are marked by ``A.''  The
first two spectra have multiple emission features and are assigned
$z_{\rm conf}=1$ redshifts.  The second spectrum from the top is an
example of how the strong emission lines in a Type 2 AGN allow
unambiguous redshifts even if the signal-to-noise is low.  The third
spectrum has only one emission line that is not an obvious artifact
and is assigned $z_{\rm conf}=2$.  The bottom spectrum is typical of
the red galaxies in our sample and with multiple absorption features
is assigned $z_{\rm conf}=1$.}
\label{fig:type2ex}
\end{figure}

\begin{figure}
\plotone{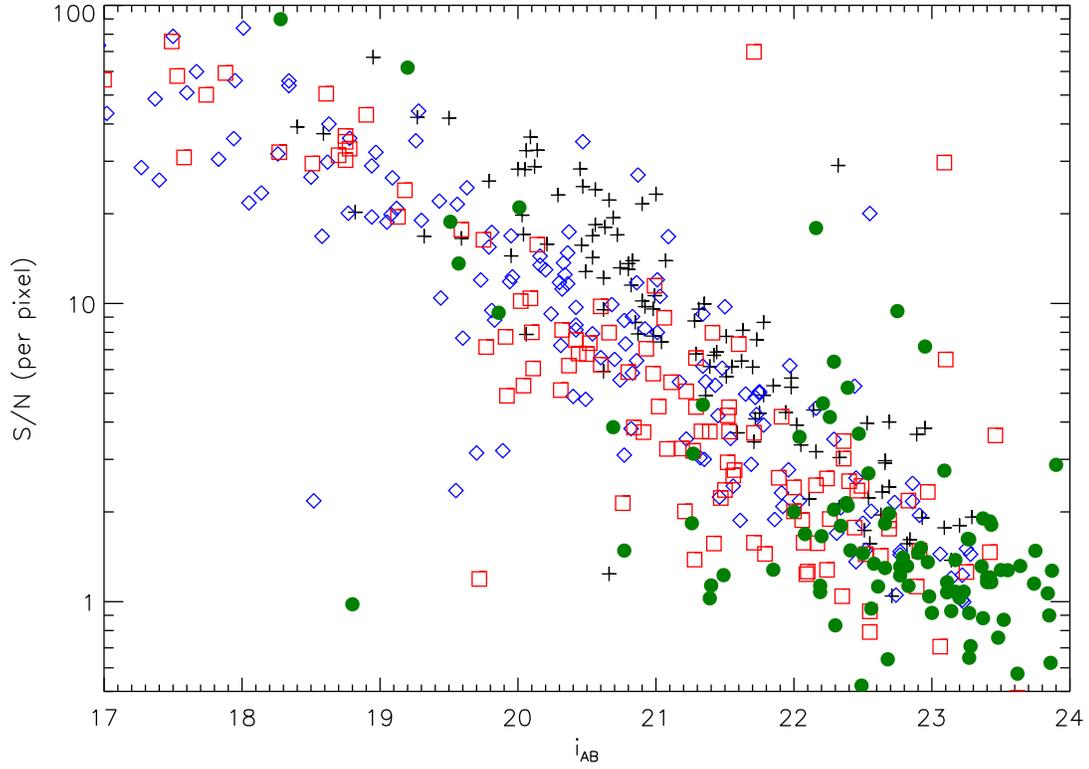}
\figcaption{The measured signal-to-noise of our spectra with their
target $i_{\rm AB}$ magnitudes.  Black crosses represent Type 1 AGN,
blue diamonds are Type 2 AGN, red squares are red galaxies, and filled
green circles are unclassified targets.  In general, signal-to-noise
increases with brighter targets.  The outlying objects had poorly
machined or misaligned slits or severe contamination from the 0th and
2nd order features of neighboring slits.  Our ability to classify
targets decreases significantly for S/N$<2.5$ and $i_{\rm AB}<22$.}
\label{fig:sni}
\end{figure}

\begin{figure}
\plotone{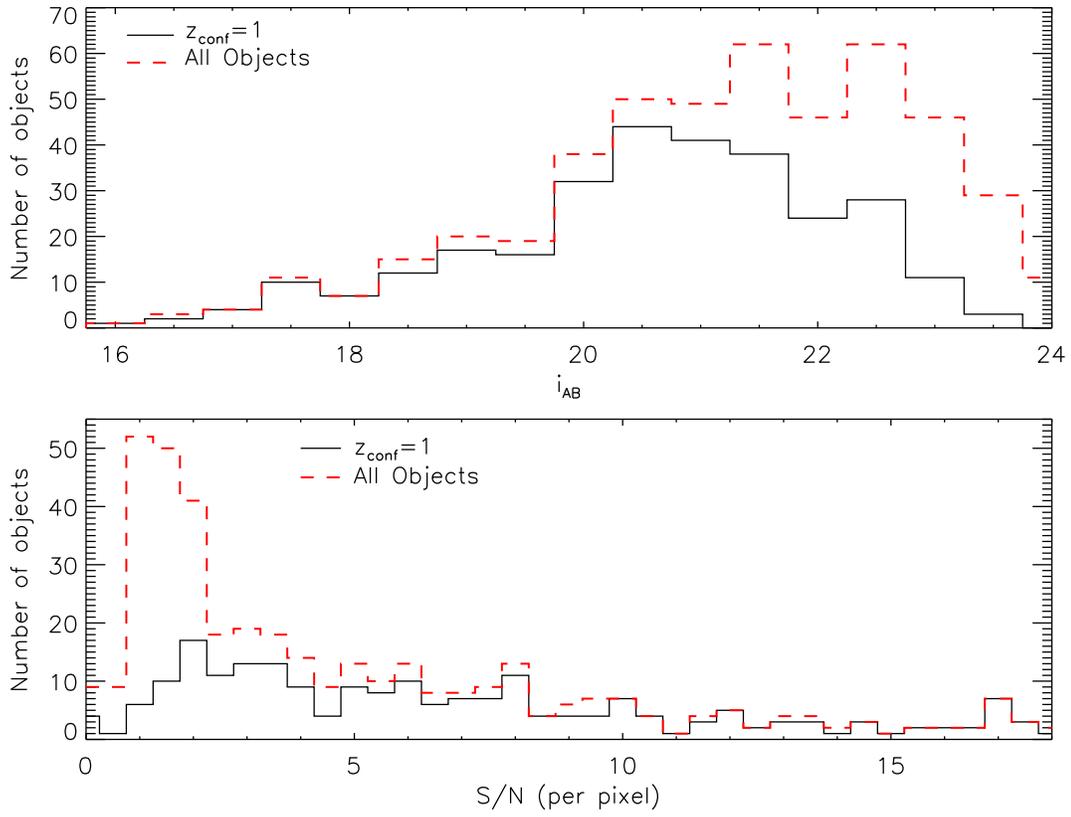}
\figcaption{The distributions of $i_{\rm AB}$ magnitude (top panel)
and signal-to-noise (bottom panel) of our sample.  The solid lines
represent objects with reliable redshifts ($z_{\rm conf}=1$) and the
dashed lines represent all objects.  A high fraction of spectra with
$S/N \gtrsim 2.5$ and $i_{\rm AB} \lesssim 22$ yield reliable
redshifts.}
\label{fig:snihist}
\end{figure}

\begin{figure}
\plotone{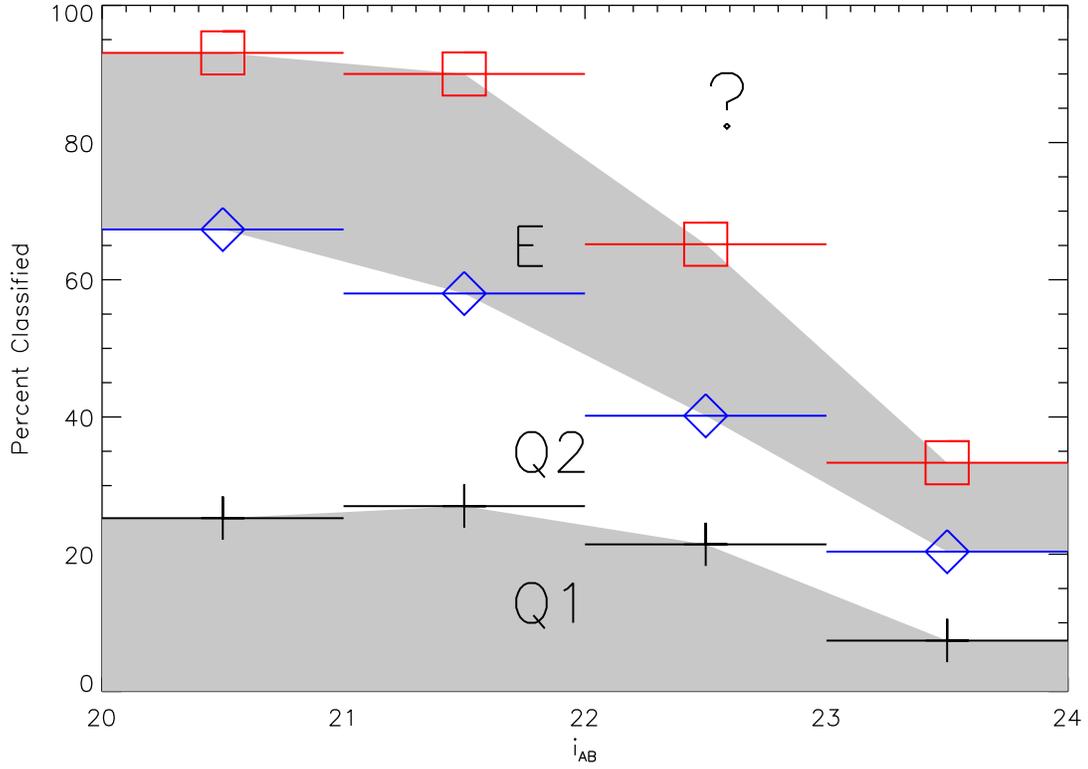}
\figcaption{The completeness of our sample with $i_{\rm AB}$
magnitude, by type.  Our completeness to type 1 AGN (the region
labeled ``Q1'') does not appear to drop off until $i_{\rm AB}>23$,
while completeness to type 2 AGN (``Q2'') and red galaxies (``E'')
drops for $i_{\rm AB}>22$.  The fraction of unclassified objects
(labeled ``?'')  increases significantly after the $21<i_{\rm AB}<22$
bin.}
\label{fig:typesi}
\end{figure}

\begin{figure}
\plotone{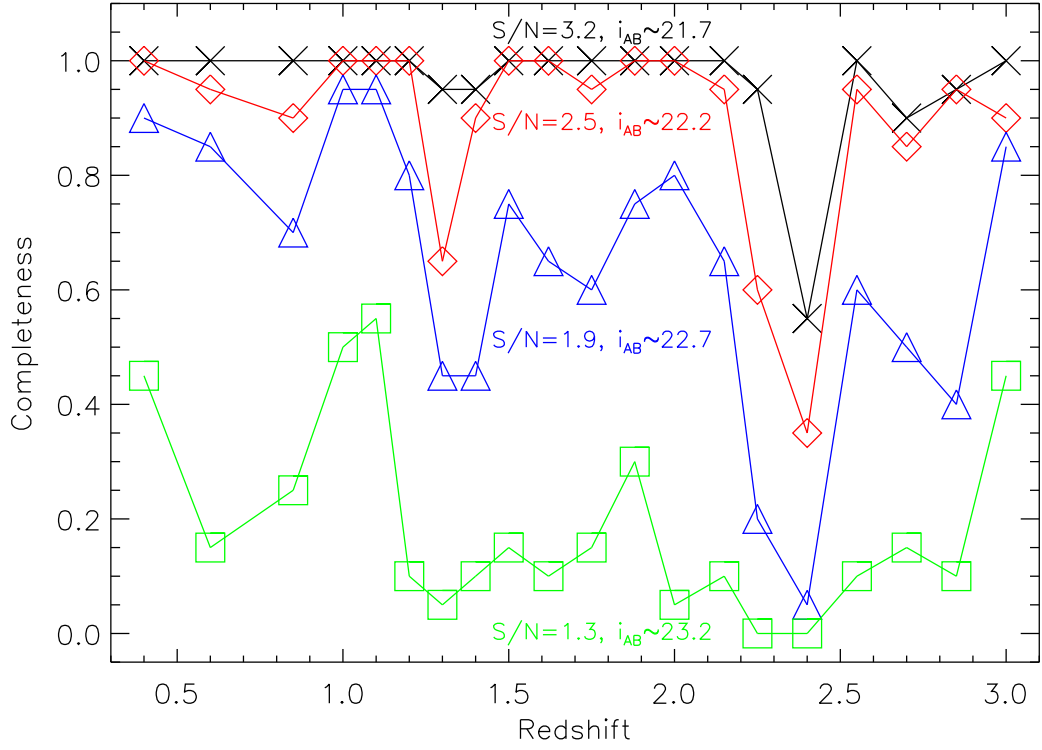}
\figcaption{Our estimated completeness to assigning $z_{\rm conf}=1$
redshifts for Type 1 AGN, as determined by our Monte Carlo
simulations.  We tested 20 iterations of Gaussian-distributed random
errors according to 4 values of signal-to-noise, over 20 redshift
bins.  Each signal-to-noise corresponds to a $i_{\rm AB}$ magnitude
according to Figure 5.  The redshift ranges of lowest completeness
correspond to the regions in which $\MgII$ and $\CIII$ are the only
lines present for our wavelength coverage.}
\label{fig:q1complete}
\end{figure}

\begin{figure}
\plotone{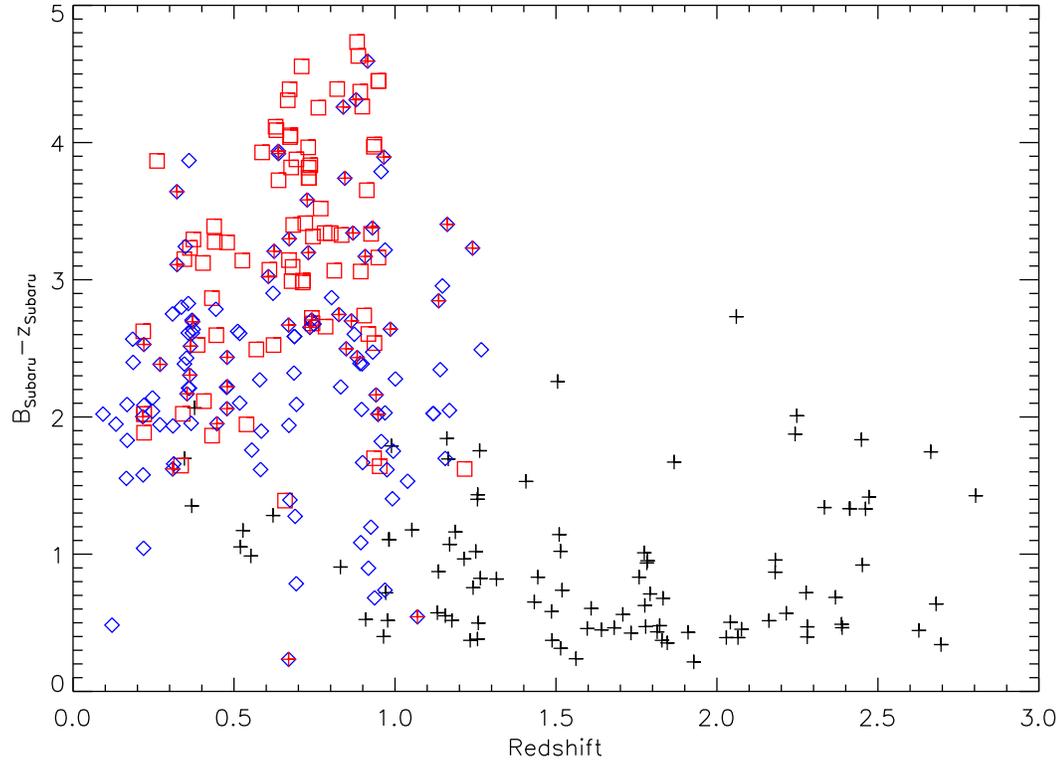}
\figcaption{The Subaru $B-z$ color of our targets with redshift.
Black crosses represent Type 1 AGN, blue diamonds are Type 2 AGN, and
red squares are red galaxies.  Hybrid ``q2e'' targets with emission
lines and red galaxy continua are blue diamonds filled with red.
Because our targets have different $B-z$ colors at all redshifts
(though especially at $z \sim 1$), we should be able to place
constraints on the unclassified targets using their $B$ and $z$
colors.}
\label{fig:}
\end{figure}

\begin{figure}
\plotone{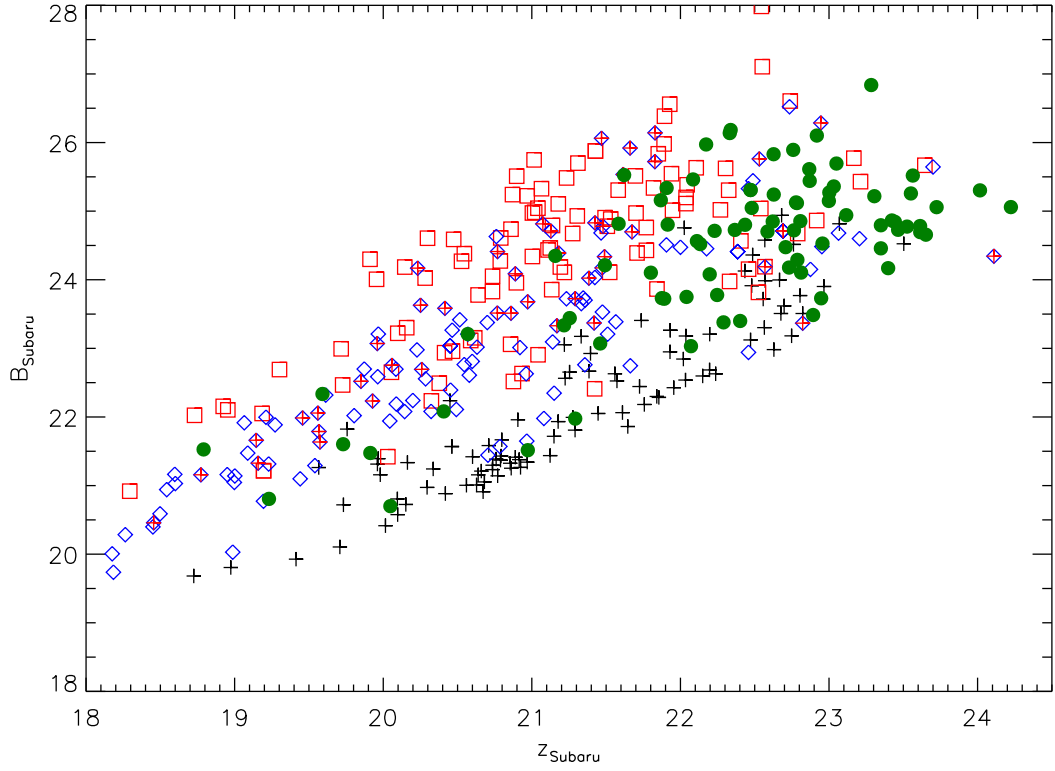}
\figcaption{The distribution of Subaru $B$ and $z$ colors for the Type
1 AGN (black crosses), Type 2 AGN (blue diamonds, with hybrid ``q2e''
objects filled red), red galaxies (red squares), and unclassified
targets (green filled circles).  For $z<23$, the colors of our
unclassified targets suggest that they are mostly red galaxies or Type
2 AGN.  We cannot constrain the $z>23$ unclassified targets as
effectively.}
\label{fig:q1complete}
\end{figure}

\begin{figure}
\plotone{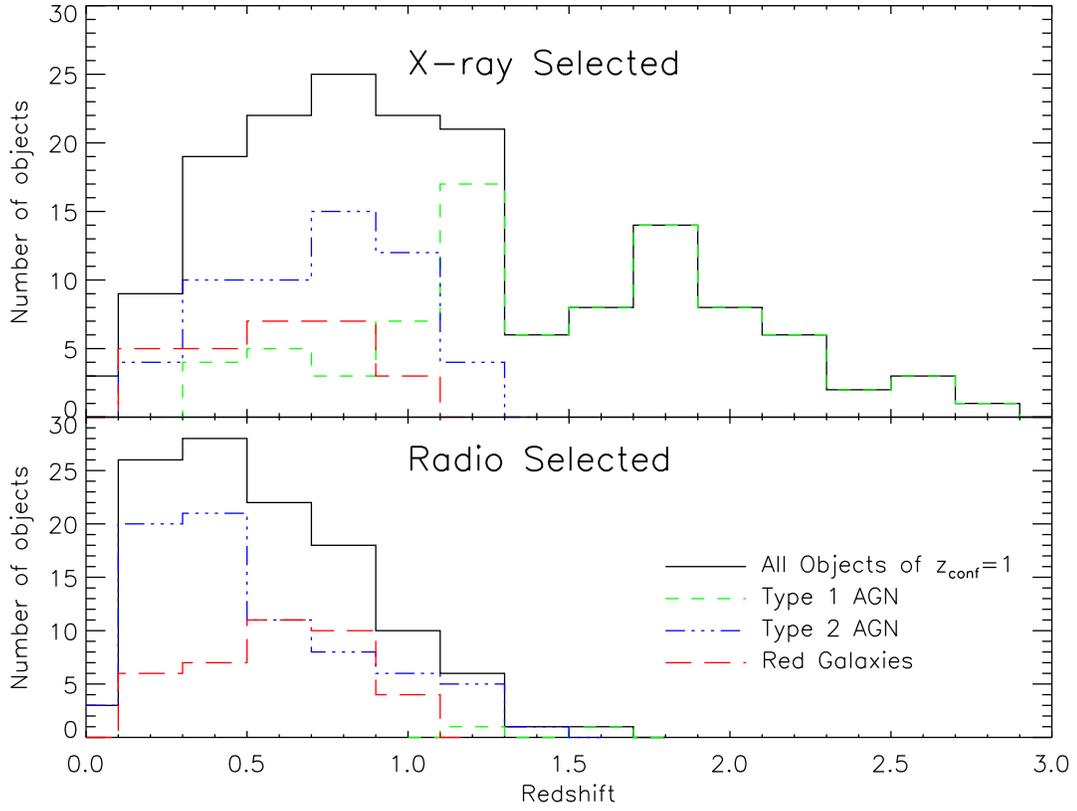} \figcaption{The distribution of redshifts for X-ray
selected and radio selected target identified in our sample.  The
solid line represents all objects of a given selection method, with
the distribution of Type 1 AGN shown by the green short-dashed line,
Type 2 AGN by the blue dash-dotted line, and red galaxies by the red
long-dashed line.  X-ray selected Type 1 AGN dominate the $z \gtrsim
1$ population of our sample.  The radio selected distribution has an
additional low redshift Type 2 population which is probably dominated
by emission line galaxies.}
\label{fig:zhist}
\end{figure}

\begin{figure}
\plotone{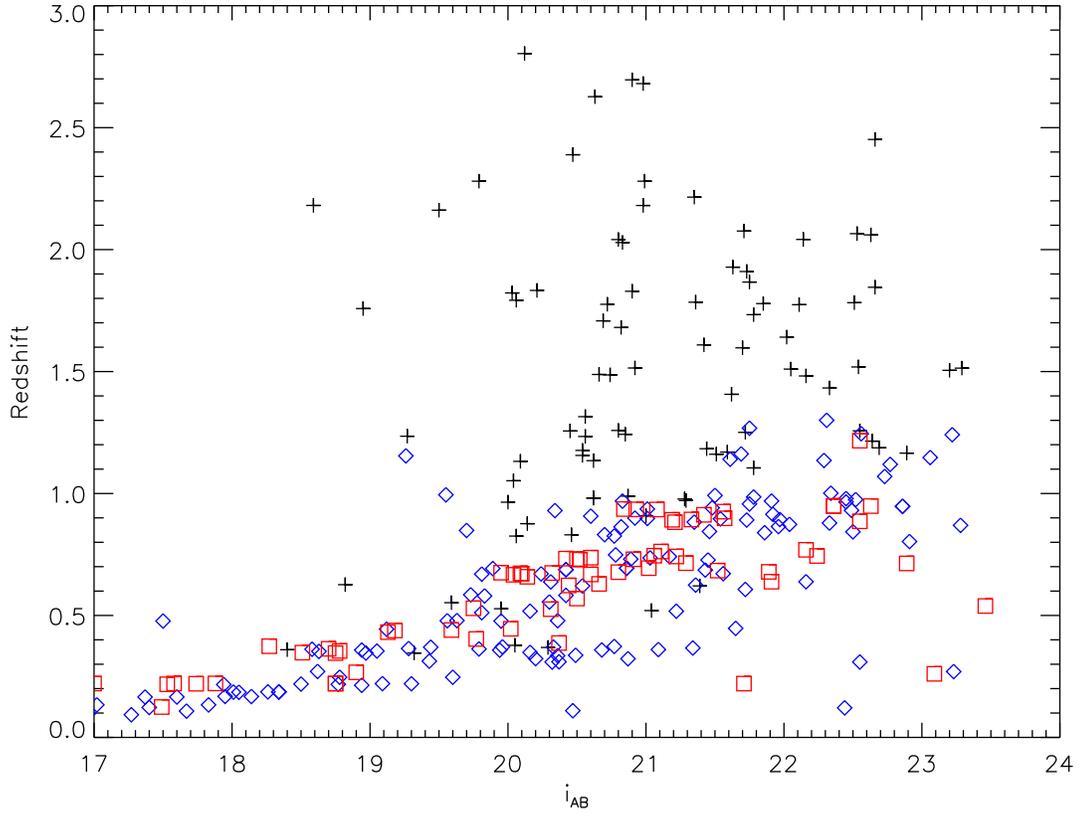}
\figcaption{The relationship between our redshifts and the $i_{\rm
AB}$ magnitudes.  Type 1 AGN are represented by crosses, Type 2 AGN by
the diamonds, and galaxies by the squares.  While the Type 2 AGN and
red galaxies follow a very similar distribution, the Type 1 AGN are on
average significantly more distant and luminous.}
\label{fig:zi}
\end{figure}

\begin{figure}
\plotone{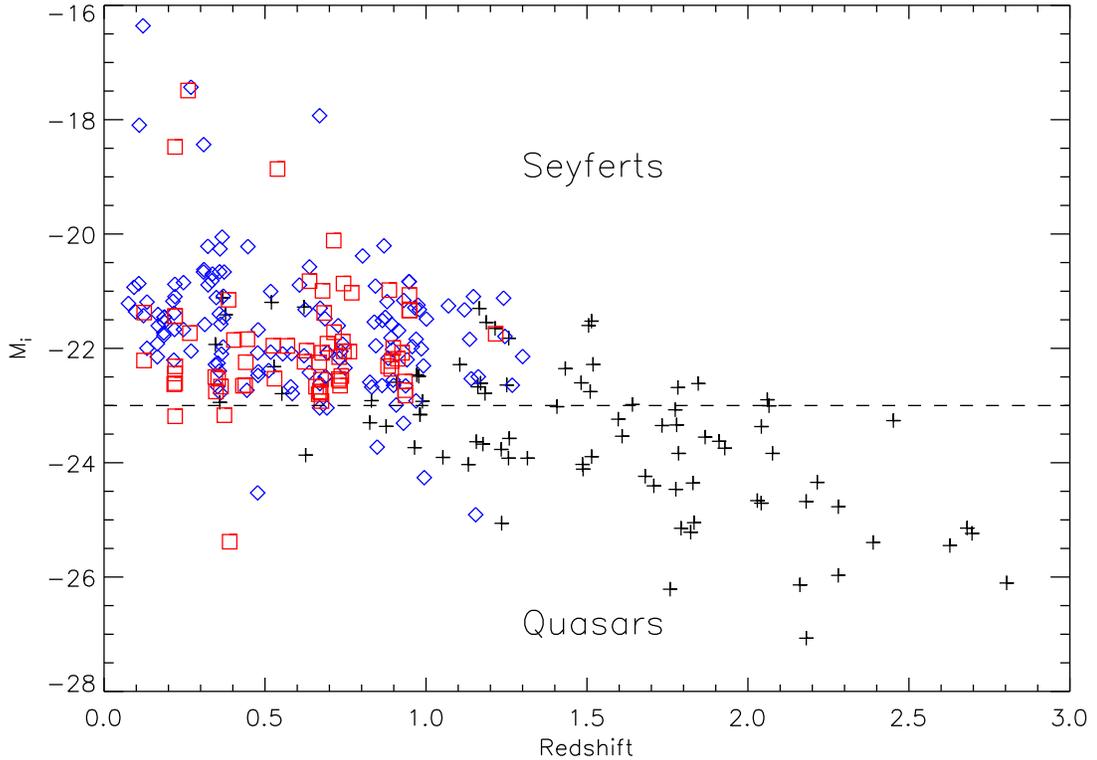}
\figcaption{The absolute $i$-magnitudes with redshifts for our
targets.  Type 1 AGN are represented by crosses, Type 2 AGN by
diamonds, and red galaxies by squares.  Our spectroscopic survey
is sensitive to a variety of Seyfert and quasar AGN for $z<1.5$.}
\label{fig:absmag}
\end{figure}

\begin{figure}
\plotone{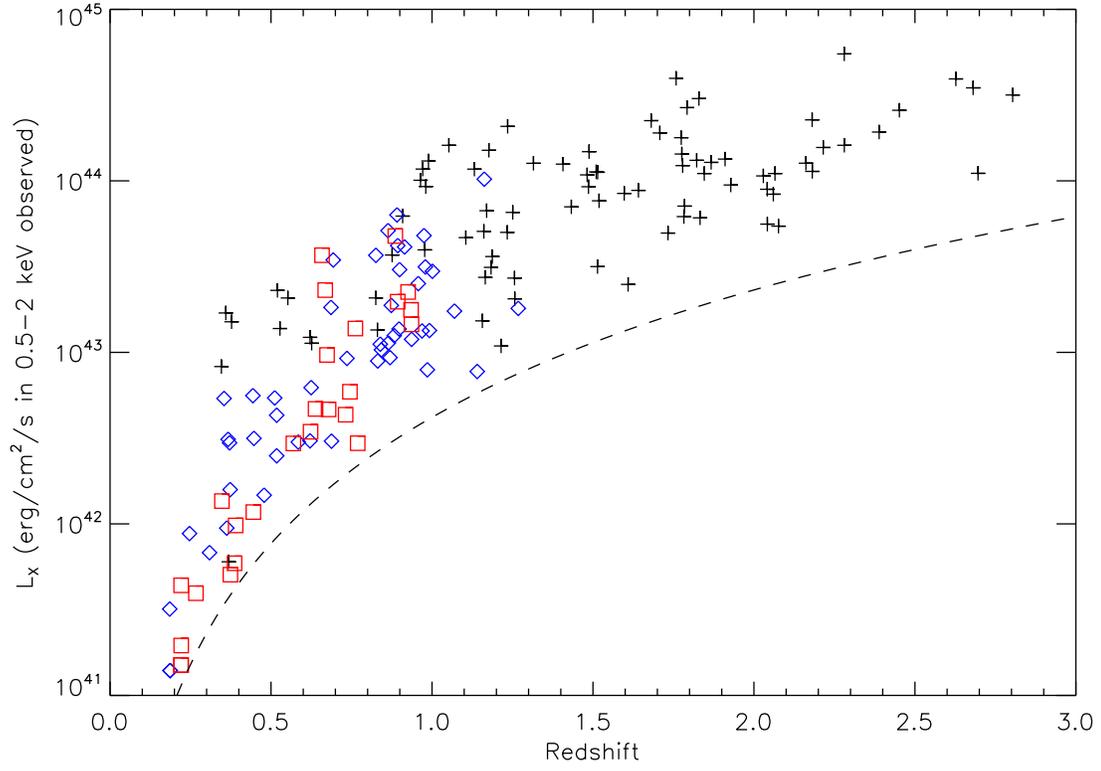}
\figcaption{The X-ray luminosities of our X-ray targets with their
redshifts.  Type 1 AGN are represented by crosses, Type 2 AGN by
diamonds, and red galaxies by squares.  The X-ray flux limit of our
targets is shown as the dashed line.  X-ray luminosities are
calculated from the X-ray flux in the observed 0.5-2 keV band.}
\label{fig:xrayfl}
\end{figure}

\end{center}

\clearpage

\begin{deluxetable}{rrrrrrr}
\tablecolumns{7}
\tablewidth{0pc}
\tablecaption{COSMOS Observation Details by Field}
\label{tbl:fields}
\tablehead{
  \colhead{Field} & 
  \colhead{$t_{\rm exp}$(Jan.)} & 
  \colhead{$t_{\rm exp}$(Feb.)} & 
  \colhead{$n_{\rm targets}$} & 
  \colhead{$n_{\rm X-ray}$} & 
  \colhead{$n_{\rm radio}$} & 
  \colhead{$n(z_{\rm conf}=1)$}}
\startdata
6 & 24840 & ... & 77 & 44 & 33 & 52 \\
7 & 12720 & 9000 & 100 & 66 & 34 & 62 \\
10 & 13200 & 21470\tablenotemark{a} & 68 & 36 & 32 & 31 \\
11 & ... & 16800 & 73 & 46 & 27 & 45 \\
12 & ... & 17160 & 65 & 38 & 27 & 32 \\
15 & ... & 13080 & 43 & 24 & 19 & 33 \\
16 & ... & 14160 & 40 & 28 & 12 & 27 \\
\enddata
\tablenotetext{a}{Field 10 was observed in ``poor seeing'' mode in
February.}
\end{deluxetable}

\begin{deluxetable}{lrrrrrlrrc}
\tablecolumns{10}
\tablewidth{600pt}
\tablecaption{COSMOS IMACS AGN catalog (page 1)}
\label{tbl:ztable}
\tablehead{
\colhead{Object Name} & \colhead{RA (J2000)} & \colhead{Dec (J2000)} & \colhead{$i_{\rm CFHT}$} & \colhead{S/N} & \colhead{$t_{\rm exp}$} & \colhead{Type} & \colhead{z} & \colhead{$\sigma_z$} & \colhead{$z_{\rm conf}$} }
\startdata
COSMOS J095859.33+022044.7 &  149.7472229 &    2.3457551 & 18.61 &  50.51 & 24840 &     e &  0.37389 &  0.00002 & 2 \\
COSMOS J095900.62+022833.3 &  149.7525635 &    2.4759071 & 19.95 &  16.87 & 24840 &    q2 &  0.47723 &  0.00007 & 1 \\
COSMOS J095900.64+021954.4 &  149.7526398 &    2.3317800 & 20.36 &  14.78 & 24840 &    q2 &  0.33492 &  0.00001 & 1 \\
COSMOS J095901.82+021449.6 &  149.7575989 &    2.2471199 & 22.39 &   2.10 & 24840 &     ? & -1.00000 & -1.00000 & ? \\
COSMOS J095902.56+022511.8 &  149.7606354 &    2.4199319 & 21.78 &   4.92 & 24840 &    q1 &  1.10490 &  0.00592\tablenotemark{b} & 1 \\
COSMOS J095902.66+022738.8 &  149.7610931 &    2.4607720 & 20.10 &   8.00 & 24840 &     e &  0.67068 &  0.00034 & 1 \\
COSMOS J095904.41+020333.8 &  149.7683563 &    2.0594010 & 21.26 &   1.83 & 13200 &     ? & -1.00000 & -1.00000 & ? \\
COSMOS J095906.97+021357.8 &  149.7790222 &    2.2327120 & 21.11 &   5.44 & 24840 &     e &  0.76203 &  0.00052 & 1 \\
COSMOS J095907.65+020820.9 &  149.7818756 &    2.1391280 & 19.05 &  18.75 & 13200 &   q2e &  0.35416 &  0.00004 & 1 \\
COSMOS J095908.23+015446.2 &  149.7842865 &    1.9128259 & 21.32 &   3.04 & 13200 &    q2 &  1.15604 &  0.00030 & 2 \\
COSMOS J095908.34+020540.7 &  149.7847443 &    2.0946369 & 17.27 &  28.58 & 13200 &    q2 &  0.09308 &  0.00004 & 1 \\
COSMOS J095908.40+020403.7 &  149.7849884 &    2.0677061 & 17.67 &  59.98 & 13200 &    q2 &  0.10792 &  0.00003 & 1 \\
COSMOS J095908.77+022315.2 &  149.7865601 &    2.3875580 & 23.06 &   0.71 & 24840 &     e &  0.91729 &  0.00432 & 2 \\
COSMOS J095909.53+021916.5 &  149.7897339 &    2.3212631 & 20.05 &  28.16 & 24840 &    q1 &  0.37753 &  0.00005 & 1 \\
COSMOS J095909.97+022017.7 &  149.7915649 &    2.3382571 & 21.41 &   7.98 & 24840 &     e &  0.43187 &  0.00156 & 2 \\
COSMOS J095910.02+020509.4 &  149.7917480 &    2.0859480 & 23.87 &   1.27 & 13200 &     ? & -1.00000 & -1.00000 & ? \\
\enddata
\tablenotetext{b}{These objects were manually assigned a redshift
error derived from the 5-pixel spectral resolution.}
\tablenotetext{c}{These $z_{\rm conf}=2$ Type 1 AGN had their
redshifts manually adjusted to be more consistent with their $u^*_{\rm
CFHT} - B_{\rm Subaru}$ and $V_{\rm Subaru} - r_{\rm Subaru}$ colors.}
\end{deluxetable}

\begin{deluxetable}{lrrrrrr}
\tablecolumns{7}
\tablewidth{0pc}
\tablecaption{Breakdown of AGN Candidates\tablenotemark{d}}
\label{tbl:efficiencies}
\tablehead{
  \colhead{} &
  \colhead{X-ray Targets} & \colhead{} &
  \colhead{Overlap} & \colhead{} &
  \colhead{Radio Targets} & \colhead{} \\
  \colhead{} &
  \colhead{Total} &
  \colhead{Example\tablenotemark{e}} &
  \colhead{Total} &
  \colhead{Example\tablenotemark{e}} &
  \colhead{Total}  &
  \colhead{Example\tablenotemark{e}} }
\startdata
Total $i_{\rm AB}<24$ sources & 800 & 58 & 150 & 10 & 700 & 43 \\
Targeted & 660 & 48 & - & - & 420 & 28 \\
Classified & 500 & 41 &  - &  - & 350 & 23 \\
Assigned $z_{\rm conf}=1$ Redshifts & 390 & 28 & - & - & 280 & 18 \\
Assigned $z_{\rm conf}=2$ Redshifts & 80 & 8 & - & - & 45 & 3 \\
\enddata
\tablenotetext{d}{We display numbers of objects in each stage of the
targeting and analysis process.  We show both the total number of
objects over all 16 IMACS pointings (the total numbers targeted,
classified, and assigned redshifts are estimated from the 7 observed
pointings) and an example of a single pointing.  The overlap columns display the number of
targets selected by both X-ray and radio emission.  Such targets are
only included in the X-ray columns for the targeted, classified, and
assigned redshift rows.}
\tablenotetext{e}{We use the number of objects in Field 11 as a typical example of the number of objects per pointing.}
\end{deluxetable}


\begin{thebibliography}{}

\bibitem[Abraham et~al.(2004)]{gemini}
  Abraham, R.~G. et~al. 2004, \aj, 127, 2455

\bibitem[Best et~al.(2005)]{best}
  Best, P.~N., Kauffmann, G., Heckman, T.~M., Ivezi\'c, Z. 2005,
  \mnras, 362, 9

\bibitem[Bigelow et~al.(1998)]{imacs}
  Bigelow, B.~C., Dressler, A.~M., Shectman, S.~A., \& Epps,
  H.~W. 1998, in Proc. SPIE Vol. 3355, p. 225-231, Optical
  Astronomical Instrumentation, Sandro D'Odorico; Ed., 225--231

\bibitem[Brusa et~al.(2006)]{brusa}
  Brusa, M. et~al. 2006, \apjs, this volume

\bibitem[Budav\'{a}ri et~al.(2001)]{q1colors}
  Budav\'{a}ri, T. et~al. 2001, \aj, 122, 1163

\bibitem[Eckart et~al.(2006)]{survey3}
  Eckart, M.~E., Stern, D., Helfand, D.~J., Harrison, F.~A., Mao,
  P.~H., \& Yost, S.~A. 2006, ApJ in press (astro-ph/0603556)

\bibitem[Eisenstein et~al.(2001)]{etemplate}
  Eisenstein, D.~J. et~al. 2001, \aj, 122, 2267

\bibitem[Fiore et~al.(2003)]{survey1}
  Fiore et~al. 2003, \aap, 409, 79

\bibitem[Glazebrook \& Bland-Hawthorn(2001)]{nodshuffle}
  Glazebrook, K. \& Bland-Hawthorn, J. 2001, \pasp, 113, 197

\bibitem[Hasinger, Miyaji, \& Schmidt(2005)]{hasinger} Hasinger, G.,
  Miyaji, T. \& Schmidt, M. 2005, \aap, 441, 417

\bibitem[Hasinger et~al.(2006)]{cosmosxray}
  Hasinger, G. et~al. 2006, \apjs, this volume

\bibitem[Impey et~al.(2006)]{cosmosagn}
  Impey, C.~D. et~al. 2006, \apjs, this volume

\bibitem[La Franca et~al.(2005)]{lafranca}
  La Franca, L. et~al. 2005, \apj, 635, 864

\bibitem[Lilly et~al.(2006)]{zcosmos}
  Lilly, S.~J. et~al. 2006, \apjs, this volume

\bibitem[Mainieri et~al.(2006)]{xrayspec}
  Mainieri, V. et~al. 2006, \apjs, this volume

\bibitem[Moustakas \& Kennicutt(2005)]{ispec}
  Moustakas, J. \& Kennicutt, R.~C. 2005, \apjs, submitted

\bibitem[Sadler et~al.(2002)]{sadler}
  Sadler, E.~M. et~al. 2002, \mnras, 329, 227

\bibitem[Schinnerer et~al.(2006)]{cosmosradio}
  Schinnerer, E. et~al. 2006, \apjs, this volume

\bibitem[Scoville et~al.(2006)]{cosmos}
  Scoville, N.~Z. et~al. 2006, \apjs, this volume

\bibitem[Silverman et~al.(2005)]{survey2}
  Silverman, J.~D. et~al. 2005, \apj, 618, 123

\bibitem[Smolcic et~al.(2006)]{radiotype2}
  Smolcic, V. 2006, in prep.

\bibitem[Steffen et~al.(2003)]{steffen}
  Steffen, A.~T., Barger, A.~J., Cowie, L.~L, Mushotzky, R.~F., \&
  Yang, Y. \apj, 596, 23

\bibitem[Sulentic et~al.(2000)]{lineshifts}
  Sulentic, J.~W., Marziani, P., \& Dultzin-Hacyan, D. 2000, \araa,
  38, 521

\bibitem[Trump et~al.(2006)]{linemeasure}
  Trump, J.~R. et~al. 2006, in prep.

\bibitem[Vanden Berk et~al.(2001)]{q1template}
  Vanden Berk, D.~E. et~al. 2001, \aj, 122, 549

\bibitem[York et~al.(2000)]{sdss}
  York, D.~G. et~al. 2000, \aj, 120, 1579

\bibitem[Zakamska et~al.(2003)]{q2template}
  Zakamska, N.~L. et~al. 2003, \aj, 126, 2125

\end{thebibliography}
\end{document}